\documentclass[
reprint,
superscriptaddress,
 amsmath,amssymb,
 aps,
 prl
]{revtex4-1}

\usepackage{graphicx}
\usepackage{dcolumn}
\usepackage{bm}
\usepackage[unicode=true]{hyperref}
\usepackage{color}
\usepackage{xcolor}
\usepackage{makecell}
\usepackage{booktabs}
\usepackage{epstopdf} 
\usepackage{cellspace}
\usepackage{hyperref}
\begin{document}

\preprint{APS/123-QED}

\title{Tunable, Flexible and Efficient Optimization of Control Pulses for Practical Qubits}

\author{Shai Machnes}
 \email{shai.machnes@gmail.com}
\affiliation{Theoretical Physics, Saarland University, 66123 Saarbr\"ucken, Germany}
\affiliation{Weizmann Institute of Science, 76100 Rehovot}

\author{Elie Ass\'emat}
\affiliation{Theoretical Physics, Saarland University, 66123 Saarbr\"ucken, Germany}

\author{David Tannor}
\affiliation{Weizmann Institute of Science, 76100 Rehovot}

\author{Frank K. Wilhelm}
\affiliation{Theoretical Physics, Saarland University, 66123 Saarbr\"ucken, Germany}

\begin{abstract}

Quantum computation places very stringent demands on gate fidelities, and experimental implementations require both the controls and the resultant dynamics to conform to hardware-specific constraints. Superconducting qubits present the additional requirement that pulses must have simple parameterizations, so they can be further calibrated in the experiment, to compensate for uncertainties in system parameters. Other quantum technologies, such as sensing, require extremely high fidelities. We present a novel, conceptually simple and easy-to-implement gradient-based optimal control technique named Gradient Optimization of Analytic conTrols (GOAT), which satisfies all the above requirements, unlike previous approaches. To demonstrate GOAT's capabilities, with emphasis on flexibility and ease of subsequent calibration, we optimize fast coherence-limited pulses for two leading superconducting qubits architectures - flux-tunable transmons and fixed-frequency transmons with tunable couplers.

\end{abstract}

\maketitle

\textit{Introduction.} The ability to drive a quantum system to a desired target in a fast and efficient manner is at the heart of emerging quantum technologies \cite{EU-Roadmap}. Finding the optimal control pulse to drive the quantum system to a desired state or to generate a desired gate has been the subject of extensive research since the first applications of Quantum Optimal Control \cite{TannorRice1985,optim6,Krotov2,Krotov1} (QOC). QOC has been applied experimentally to photochemical reactions \cite{optim-application1}, where progress continues to this day \cite{optim-application2}. It has also been applied to nuclear magnetic resonance \cite{optim6,GRAPE,optim4}, with applications in medical imaging and spectroscopy. New experimental methods to control quantum systems have led to increased interest in QOC, which has been applied to processes as diverse as high harmonic generation \cite{optim-application3}, control of energy flow in biomolecules \cite{optim-application4}, attosecond physics \cite{atto1} and quantum computing \cite{optim-quan-comp1,optim-quan-comp2}. Superconducting qubits raise additional challenges to QOC, as fabrication variability implies pulses are often optimized in simulation using a somewhat inaccurate model of the system. To achieve high fidelity for such mischaracterized systems, a second, in situ optimization of pulse parameters is needed (also known as calibration or tune up). The latter requires simple functional forms of the pulses.

The paper is organized as follows: First, we present criteria for an ideal QOC algorithm, with specific emphasis on unique needs of superconducting qubits. Second, we formally define the optimal control task and present a new and extremely simple QOC algorithm, GOAT, which we believe is the first to simultaneously satisfy all criteria. Third, we apply GOAT to superconducting qubit systems - the flux tunable coupler \cite{IBM-tunable-coupler} and frequency-tunable qubits \cite{Xmon-PRL}, producing simple, realistic, bandwidth-limited pulses, which implement coherence-limited two-qubit gates, significantly shorter than has currently been achieved in experiments.


\textit{Requirements of QOC.} A practical QOC method ideally meets the following three criteria:

(i) \emph{Flexibility}. A QOC method must be flexible enough to accurately model the experimental system, including all control constraints, transfer functions, etc. Moreover, it must be flexible enough to utilize any control ansatz, so a simple control pulse which suits the system in question can be found. Manufacturing solid-state qubits results in variations between samples and, therefore, Hamiltonians which are not known with the required precision (this is true to a lesser extent in most quantum systems). To achieve highly accurate controls, a second closed-loop in-experiment optimization is required, to calibrate the drive shape parameters to the specific sample \cite{Xmon-leakage,PRA-2013-rosi}. Therefore, a good QOC method produces pulses described by only a few parameters, so that subsequent calibration is feasible. The calibration can then be performed using methods such as Ad-HOC \cite{optim2} or ORBIT \cite{ORBIT}. This is a reason for the popularity of the DRAG method \cite{DRAG}, a fully analytic method to design few-parameter pulses avoiding unwanted transitions. Unfortunately, DRAG cannot be extended to arbitrary constraints.

(ii) \emph{Numerical accuracy}. A QOC method must be numerically accurate. With technologies aimed at quantum computation, one must achieve error rates below an error-correction threshold \cite{threshold-theorem-Aharonov-benOr,Proc-R-Soc-A-1998-Preskill,quan-comp-accuracy2,quan-comp-accuracy1}. Surface codes, for example, \cite{Quant-Info-Comput-2011-Fowler} require gate infidelities of $10^{-4}$ to limit overhead \cite{ladd2010quantum}. Ion trap architectures \cite{quan-comp18, Gaebler2016} and quantum circuits based on Josephson junctions, are approaching this threshold \cite{quan-comp16}. Other applications, such as sensing and metrology, require extremely high fidelities. A good QOC method must not make any approximations which degrade numerical accuracy, as this may lead to false fidelity estimates.

(iii) \emph{Speed} As a practical tool, a good QOC method should be fast, even when the number of parameters is in the double-digits, and preferably easily parallelizable. There are multiple factors affecting overall computation effort. Primarily, the effort depends on the number of iterations required to reach the desired fidelity, implying that the search of parameter space must be gradient driven. In addition, several additional factors are of importance, such as the processing done at each iteration. For example, when using piecewise-constant controls, the effort scales linearly with the number of time slices and can be very inefficient.


\textit{The QOC task.} Assume a system whose dynamics is described by a drift Hamiltonian $H_{0}$ and a set of control Hamiltonians $H_{k}$. The total Hamiltonian reads
\begin{equation}
H\left(\bar{\alpha},t\right)=H_{0}+\sum_{k=1}^{C}c_{k}\left(\bar{\alpha},t\right)H_{k}\,,
\label{eq:Ham_tot}
\end{equation}
where $c_k$ are the control functions, characterized by a set of parameters $\bar{\alpha}$. For example, the controls may be a superposition of Gaussian pulses,
\begin{equation}
c_{k}\left(\bar{\alpha},t\right)=\sum_{j=1}^{m}A_{k,j}\exp\left(-(t-\tau_{k,j})^2/\sigma_{k,j}^2\right),
\label{eq:Fourier-controls}
\end{equation}
with
\begin{equation}
\bar{\alpha}=\left\{ A_{k,j},\tau_{k,j},\sigma_{k,j}\right\} _{k=1\ldots C,j=1\ldots m}.
\label{eq:p-bar}
\end{equation}
The choice of control ansatz is governed by two considerations: constraints and sparsity. The former relates to the ease with which experimental control constraints can be modeled by the ansatz (e.g. if the control is bandwidth limited, a Fourier representation is natural). The latter relates to producing pulses which are described by only a few parameters (and, therefore, easily calibrated). In the examples below we have found Fourier and Erf parameterizations to low parameter counts pulses.

For our purpose, the goal function to minimize is defined as the projective $SU$ distance (infidelity) between the desired gate,  $U_{\textrm{goal}}$, and the implemented gate, $U\left(T\right)$, \cite{PalaoKosloff2002} (also \cite{TESCH2001633})
\begin{equation}
g\left(\bar{\alpha}\right):=1-\tfrac{1}{\textrm{dim}\left(U\right)}\left|\textrm{Tr}\left(U_{\textrm{goal}}^{\dagger}U\left(T\right)\right)\right|\,,\label{eq:goal}
\end{equation}
where $U\left(t\right)$ is the time ordered ($\mathbb{T}$) evolution operator
\begin{equation}
U\left(\bar{\alpha},T\right)=\mathbb{T}\exp\left(\int_{0}^{T}-\frac{i}{\hbar}H\left(\bar{\alpha},t\right)dt\right).\label{eq:Upt}
\end{equation}

\textit{GOAT.} We now present a novel QOC algorithm which uniquely meets all criteria described above. GOAT's ability to use any control ansatz makes it feasible to find drive shapes described by a small number of parameters, suitable for calibration.

A gradient-based optimal control algorithm requires two ingredients: an efficient computation of $\partial_{\bar{\alpha}}g\left(\bar{\alpha}\right)$ and a gradient-based search method over parameter space. GOAT presents a novel method for the former, while using any standard algorithm for the latter. Consider the gradient of the goal function Eq. (\ref{eq:goal}) with respect to $\bar{\alpha}$,
\begin{equation}\label{eq:goal-func-grad}
\partial_{\bar{\alpha}}g\left(\bar{\alpha}\right) =-\textrm{Re}\left(\frac{g^{*}}{\left|g\right|}\frac{1}{\textrm{dim}\left(U\right)}\textrm{Tr}\left(U_{\textrm{goal}}^{\dagger}\partial_{\bar{\alpha}}U\left(\bar{\alpha},T\right)\right)\right)\,.\\
\end{equation}
Neither $U\left(\bar{\alpha},T\right)$ nor $\partial_{\bar{\alpha}}U\left(\bar{\alpha},T\right)$ can be described by closed form expressions. $U$ evolves under the equation of motion $\partial_t U\left(\bar{\alpha},t\right) = -\frac{i}{\hbar}H\left(\bar{\alpha},t\right) U\left(\bar{\alpha},t\right)$. By taking the derivative of the $U$ e.o.m. with respect to $\bar{\alpha}$ and swapping derivation order, we arrive at a coupled system of e.o.m.-s for the propagator and its gradient,
\begin{equation}
\partial_{t}\left(\begin{array}{c}
U\\
\partial_{\bar{\alpha}}U
\end{array}\right)=-\frac{i}{\hbar}\left(\begin{array}{cc}
H & 0\\
\partial_{\bar{\alpha}}H & H
\end{array}\right)\left(\begin{array}{c}
U\\
\partial_{\bar{\alpha}}U
\end{array}\right).\label{eq:joint-eom}
\end{equation}
As $\bar{\alpha}$ is a vector, $\partial_{\bar{\alpha}}U$ represents multiple equations of motion, one for each component of $\bar{\alpha}$. $\partial_{\bar{\alpha}}H$ is computed using the chain rule and eqs. (\ref{eq:Ham_tot},\ref{eq:Fourier-controls}). We note that Eq. (\ref{eq:joint-eom}) was first presented in \cite{gradient-comput1}, but was not used for QOC.

GOAT optimization proceeds as follows: Starting at some initial $\bar{\alpha}$ (random or educated guess), initiate a gradient driven search (e.g. L-BFGS \cite{L-BFGS}) to minimize Eq. (\ref{eq:goal}). The search algorithm iterates, requesting evaluation of eqs. (\ref{eq:goal},\ref{eq:goal-func-grad}) at various values of $\bar{\alpha}$, and will terminate when the requested infidelity is reached or it fails to improve $g$ further. Evaluation of $g\left(\bar{\alpha}\right)$, $\partial_{\bar{\alpha}}g\left(\bar{\alpha}\right)$ requires the values of $U\left(\bar{\alpha},T\right)$ and $\partial_{\bar{\alpha}}U\left(\bar{\alpha},T\right)$. These are computed by numerical forward integration of Eq. \ref{eq:joint-eom}, by any mechanism for ODE integration that is accurate and efficient for time-dependent Hamiltonians, such as adaptive Runge-Kutta. Initial conditions are $U\left(t=0\right)=\mathcal{I}$ and $\partial_{\bar{\alpha}}U\left(t=0\right)=0$. Note that no back propagation is required.

Experimental constraints can be easily accommodated in GOAT by mapping the optimization from an unconstrained space to a constrained subspace, and computing the gradient of the goal function using the chain rule. For example, $\bar{\alpha}$ components may be constrained by applying bounding functions, e.g. $\alpha_k  \longrightarrow  \frac{1}{2}\left(v_{\text{max}} - v_{\text{min}}\right)\sin\left(\bar{\alpha}_k\right) + \frac{1}{2}\left(v_{\text{max}} + v_{\text{min}}\right)$ which imposes $\alpha_k\in\left[v_{\text{min}} \ldots v_{\text{max}}\right]$. Amplitude constraints and a smooth start and finish of the control pulse can be enforced by passing the controls through a window function which constrains them to a time-dependent envelope. Gradients for $\partial_{\bar{\alpha}}H$ flow via the chain rule. See \hyperref[sec:Appendix-C]{Appendix C} for a fully worked out example.

Application of GOAT to state transfer, open systems and super-operator generation are possible by replacing Eq. (\ref{eq:joint-eom}) with a derivative of the suitable equation of motion with respect to $\bar{\alpha}$, see \cite{Shai-PRA}. Filters may be modeled by including the filter's internal state e.o.m., alongside Eq. (\ref{eq:joint-eom}), when deriving by $\bar{\alpha}$. See also \cite{hincks2015controlling}. Equations (\ref{eq:goal-func-grad}) and (\ref{eq:joint-eom}) can be modified to provide second-order gradient information, allowing Hessian-driven search, such as Newton-Raphson (see \cite{goodwin2015modified}).


\textit{Comparison with current algorithms.} Examination of the prevailing QOC methods reveals none meet all three criteria for an ideal QOC method:

One class of QOC methods is based on gradient-free optimization of the parameters:
sample $g\left(\bar{\alpha}\right)$ at several $\bar{\alpha}$-s, deduce one or more new $\bar{\alpha}$-s for which $g$ is expected to be lower, and repeat.
This approach is simple and flexible, and is the only possible procedure for closed-loop calibration.
However, it converges very slowly compared to gradient driven optimization, particularly when optimizing high-dimensional parameter spaces.
For example, the Nelder-Mead optimization algorithm  \cite{Nelder-Mead}, at the basis of the CRAB and dCRAB methods \cite{CRAB,dCRAB}, grows excessively slow when the number of parameters approaches $10$ \cite{Nelder-Mead-slow-dim} (dCRAB can be viewed as successive CRAB searches of alternate subspaces).
Other methods, such as CMA-ES \cite{CMA-ES}, genetic algorithms or Simultaneous Perturbation Method \cite{spall1992multivariate}, are somewhat better at handling large parameter spaces, but are still slow to converge compared to gradient driven methods. When the gradient of the goal function with respect to the parameterization, $\partial_{\bar{\alpha}}g\left(\bar{\alpha}\right)$, can be computed efficiently, gradient driven optimization algorithms outperform gradient-free methods by orders of magnitude (see \hyperref[sec:Appendix-A]{Appendix A} and references within \cite{nocedal2006no,L-BFGS}). Thus, gradient-free methods fail criterion (iii). In contrast, GOAT is gradient driven, and can utilize any gradient driven search algorithm (including second order methods, such as Newton), and, therefore,  converges quickly, satisfying criterion (iii).

A second class of QOC methods, such as Krotov \cite{krotov1983iteration,Tannor1992,Koch-Krotov-Main} and GRAPE \cite{GRAPE}, derive from a variational formulation of the QOC task \cite{pontryagin1986mathematical}, where the Schr\"odinger equation is imposed as a constraint.
This necessitates propagating an adjoint operator backward in time from the goal gate, acting as a Lagrange multiplier.
The update rules for the control fields in both the Krotov and GRAPE methods are defined in terms of time-local expressions, implying a piecewise constant (PWC) control ansatz.
This presents two types of problems.
First, the PWC ansatz is incompatible with the low parameter counts needed for subsequent pulse calibration.
Moreover, it does not lend itself to the imposition of control constraints, such as bandwidth, nor the freedom to choose a control ansatz.
And while workarounds have been found for both GRAPE and Krotov (\cite{optim-Skinner,optim-quan-comp1} and \cite{palao2013steering,goerz2015hybrid} respectively), these are nontrivial to implement.
Further, the variational formulation necessitates a nontrivial rederivation of the control update rule whenever a change is made to the goal functional (e.g. a new penalty term).
Thus, GRAPE and Krotov fail criterion (i) flexibility.
In contrast, GOAT, which does not derive from the variational formulation, does not require back propagation. It can easily adapt to new goal functions, utilize an arbitrary control ansatz to produce a simple calibration of pulses, and impose a wide range of constraints, meeting criterion (i).
Second, a PWC approximation of smooth low-bandwidth controls, introduces significant numerical inaccuracies in the control fields, and subsequently in the simulated dynamics, as demonstrated in \hyperref[sec:Appendix-B]{Appendix B} and references within \cite{PWC-is-crap-1,PWC-is-crap-2}.
Thus, GRAPE and Krotov fail criterion (ii) numerical accuracy. GOAT, which allows arbitrary piecewise-\emph{continuous} controls, does not suffer from this problem.

While somewhat subjective, we believe GOAT to be uniquely easy to understand and code, with a simple mathematical structure (application of the chain rule with modified equations of motion) and compatible with off-the-shelf tools for gradient driven search and ODE propagation.


\emph{Flux tunable coupler.} We consider the flux tunable coupler presented in \cite{IBM-tunable-coupler}, where a $200$ns iSWAP gate was implemented with a fidelity of $0.982$. This system consists of two transmon qubits coupled to a tunable bus resonator. The Hamiltonian reads
\begin{eqnarray}
H = \sum_{k=1}^{2} & \, &  \omega_k a_k^\dagger a_k + \omega_{\textrm{TB}}(\Phi)b^\dagger b\\
                   &   & + g_k(a_k^\dagger b + b^\dagger a_k) - \alpha_k \left| 2 \right\rangle \left\langle 2\right|_k\,,\nonumber
\end{eqnarray}
where $\omega_k$ are the frequencies of the qubits, $\omega_{\textrm{TB}}(\Phi)$ is the flux-dependent frequency of the tunable bus, $g_k$ are the coupling qubit-resonator couplings, $\alpha_k$ are the qubit anharmonicities, and $a_k$, $b$ denote the annihilation operators of the qubits and the tunable bus, respectively. The dependence of the bus frequency on the modulated flux is
\begin{equation}
\omega_{\textrm{TB}}(\Phi) = \omega_{\textrm{TB},0}\sqrt[]{|\cos(\pi \Phi/\Phi_0)|}
\label{eq:IBM-flux-dependence}
\end{equation}
\begin{equation}
\Phi = \Theta + \delta\left(t\right) \cos(\omega_\Phi t)
\end{equation}
where $\Phi_0$ is the flux quantum, $\delta(t)$ is the controlled amplitude of the flux modulation and $\omega_\Phi$ is tuned to resonantly couple $\left|01 \right\rangle$ to $\left|10 \right\rangle$.

\begin{figure}[htbp]
\begin{center}
\includegraphics[scale=0.28]{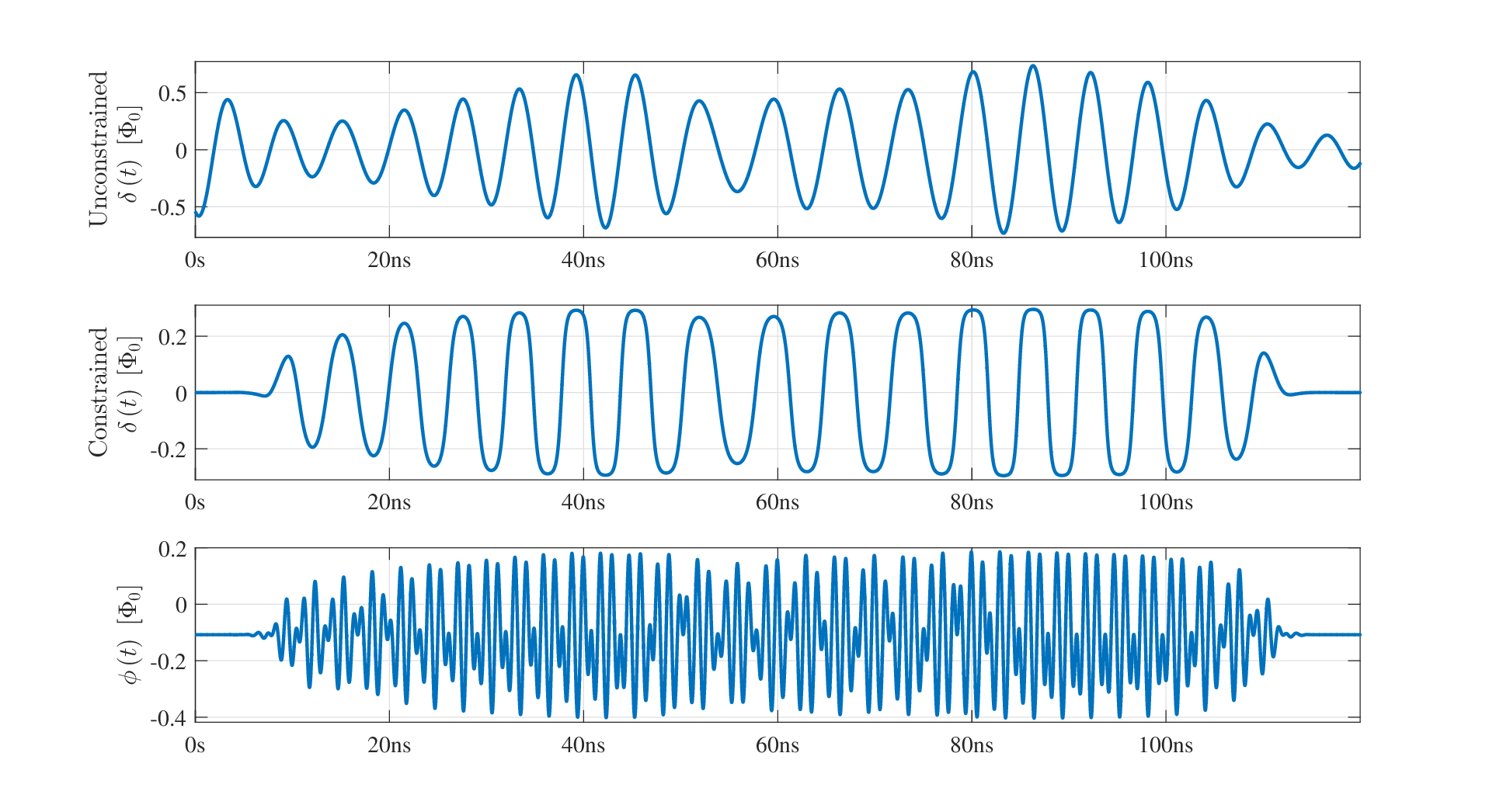}
\end{center}
\caption{Optimized flux modulation $\delta\left(t\right)$ generating an iSWAP in the flux-tunable coupler of Eq. (\ref{eq:IBM-flux-dependence}). The infidelity is $10^{-12}$ in a fully coherent model and a noise-limited $99.5\%$ with $T_1=T_2=40{\mu}$s and thermalization to $25$mK \cite{kandala2017hardware}. (top) Unconstrained Fourier pulse, sum of $6$ frequency components. (middle) Constraints applied using sigmoids, in both time (pulse starts and ends smoothly at zero) and amplitude (limited to $0.3 \Phi_0$). (bottom) Output of waveform generator, including carrier and DC bias. See \hyperref[sec:Appendix-C]{Appendix C} for further details.}
\label{fig-iSWAP-IBM}
\end{figure}

Optimizing pulses for this system use GOAT's flexibility in several ways. First, as the experiment allows for correction of the single qubit Z rotations in software, we used a modified goal function that is independent of such rotations. See \hyperref[sec:Appendix-D]{Appendix D} for further details. Then, contrary to \cite{IBM-tunable-coupler}, we do not use either the dispersive regime approximation, nor the rotating wave approximation. Rather, the optimization includes the carrier frequency (see Fig. \ref{fig-iSWAP-IBM}). This implies a more accurate simulation, and optimal pulses which require less calibration. Finally, we iteratively reduced the complexity of the control pulse: starting with a Fourier parameterization with tens of components, we successively pruned the low amplitude components and re-optimized, to reach 6 frequency components for a $100$ns iSWAP. This yielded an infidelity of $10^{-12}$ in the fully coherent model. Assuming Markovian noise with $T_1=T_2=40{\mu}$s and thermalization to $25$mK \cite{kandala2017hardware}, the pulse achieves the decoherence limit of $99.5$\% fidelity. It is possible to improve the fidelity further by two approaches. First, by shortening the pulse. This, however, requires more complex pulse shapes, which are harder to calibrate. Second, one can utilize non-Markovian features of the noise, with dynamic decoupling or spin-echo dynamics.

\textit{Calibration of simple cZ-pulses for flux-tunable qubits.} We consider a line of frequency-tunable qubits \cite{Xmon-PRL} with nearest-neighbor couplings. Define
\begin{equation}
a = \begin{pmatrix}
0 & 1 & 0 \\
0 & 0 & \sqrt{2} \\
0 & 0 & 0 \\
\end{pmatrix} \quad X=a^\dagger + a, \, Y=i\left(a^\dagger - a\right), N= a^{\dagger} a \,.
\label{Xmon-def}
\end{equation}
\begin{equation}
H = \sum_{k=1}^{2} \epsilon_k I + q_k N_k + \eta_k \left| 2\right\rangle \left\langle 2\right| + g_{k,k+1} Y_k Y_{k+1} \,.
\label{Xmon-ham}
\end{equation}
Parameters of the bare Hamiltonian are taken from recent experiments \cite{Xmon-Nature}. We implement the control-Z (CZ) gate in a two tunable-qubits system using one $z$ control per qubit, with the drive shape parameterized by error functions ($erf$). After optimization, we obtain a control signal described by only 16 parameters. Each control is a sum of two terms
\begin{align}
a_k(t) = & \frac{A_k}{4}\left(1+ \textrm{Erf}\left(\sqrt{\pi}\frac{s_k}{A_k}(t-t_{1i})\right)\right)\nonumber \\
& \times \textrm{Erfc}\left(\sqrt{\pi}\frac{s_k}{A_k}(t-t_{2i})\right).
\label{explicit-optimal-solution}
\end{align}
A sigmoid function envelops the total control amplitude, enforcing the limit. Optimization achieves a 30ns CZ gate with infidelity of $10^{-13}$ when neglecting Markovian effects, and a coherence-limited pulse when incoherent processes are introduced. This is 25\% faster than the CZ presented in the Supplemental information of \cite{quan-comp17}.

\begin{figure}[htbp]
\begin{center}
\includegraphics[scale=0.8]{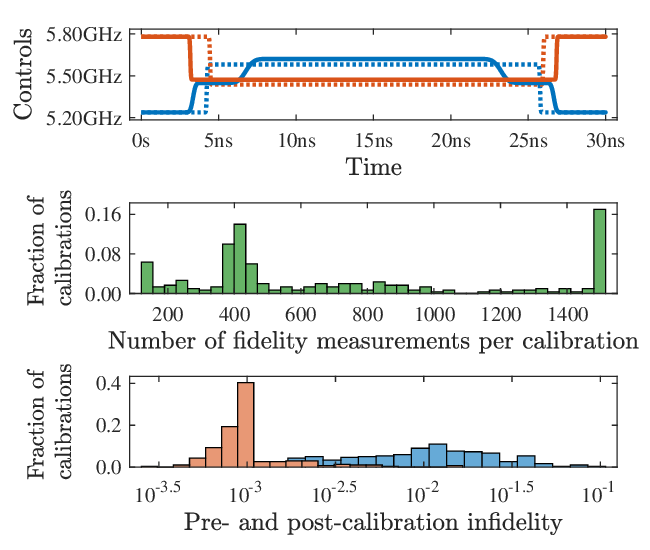}
\end{center}
\caption{Simulation of 300 GOAT pulse calibrations, assuming inaccurate characterization of system parameters. Fractional error in the coupling $g$ and anharmonicity $\eta$ is taken to be normal-distributed, with standard deviation of 3\%. Top: initial (dotted) and final (solid) drive shape for a two tunable-qubits cZ-gate parameterized with error functions using only $16$ parameters. Red and blue lines depict the two qubit controls, $a_1\left(t\right)$ and $a_2\left(t\right)$  of Eq. (\ref{explicit-optimal-solution}). Middle: distribution of the number of fidelity measurements needed to calibrate the pulse using the quasi-Newton algorithm. The low pulse parameter count implies less than 1500 fidelity measurements are needed in most simulated systems. Bottom: distribution of pre- (blue) and post- (red) calibration infidelity. Pulses have been calibrated up to the coherence limit imposed by Markovian processes, $10^{-3}$. }
\label{fig-CZ}
\end{figure}

Consider the more realistic scenario where some physical parameters are known only within a few percent. We simulated calibration by closed-loop optimization of the optimal drive parameters, to compensate for Hamiltonian mischaracterizations. The procedure is very similar to Ad-HOC \cite{optim2}, but requires an order of magnitude fewer fidelity measurements, thanks to the low number of parameters characterizing the drive shape. The choice of gradient-free algorithm (quasi-Newton instead of Nelder-Mead) reduced the number of measurements by an additional 10\%. In our simulation, we considered the coupling $g$ and the anharmonicities $\eta$ as random variables with Gaussian statistics and standard deviation of 3\%. For each instance we applied the gradient-free algorithm for closed-loop calibration, counting goal function calls, which correspond to an experimental measurement of gate fidelity. The results of the simulated calibration are shown in Fig. \ref{fig-CZ}. The key result is that for most random instances, only a few hundred fidelity measurements are required for calibration, Fig. \ref{fig-CZ}b, to reach the final fidelity, Fig. \ref{fig-CZ}c. Ad-HOC could not be used efficiently in experiments because the required number of fidelity measurements was on the order of thousands. GOAT solves this issue by reducing the number of parameters, which in turn decreases the number of fidelity measurements to the point where the Ad-HOC approach becomes a viable experimental option. The target infidelity was set at $10^{-3}$, below the coherence limit reachable for these gate durations \cite{kandala2017hardware}. Conversely, around 20\% of cases did not reach the $10^{-3}$ threshold, leading to the peak at 1500 fidelity measurements in the middle plot. Improvements to the gradient-free search algorithm would allow detecting this behavior early.


\textit{Discussion.} We have presented the criteria for good QOC algorithms: flexibility, accuracy and speed. Flexibility requires the algorithm to faithfully model the experiment, including control capabilities. This ensures resultant pulses, once implemented, will produce the desired dynamics. Moreover, it must allow the use of any control ansatz, so that one may produce pulses described by only a few parameters, enabling realistic calibration, to bridge the gap between experimentx and simulated model. Accuracy insures we do not downgrade model accuracy due to numerical issues. Speed makes the QOC process practical, allowing optimization of more accurate models.

We presented a novel optimal control algorithm, GOAT, based on equations of motion for the gradient of the propagator with respect to the drive parameters. Surveying prevailing QOC methods, we conclude that GOAT is the only QOC method to satisfy all three criteria. We demonstrated GOAT's flexibility by optimizing pulses for two different systems, using two different ansatz ($erf$ and Fourier), applying amplitude and bandwidth constraints in both cases, achieving fidelities significantly beyond current state-of-the-art. Further, we have shown the feasibility of calibrating GOAT pulses, enabled by the small number of parameters which describe them. GOAT's mathematical formulation is straightforward, and does not require backward-propagating adjoint states or the calculus of variations. It is also extremely simple to implement. GOAT does not rely on a PWC representation of bandwidth-limited controls, and therefore its accuracy is only limited by how precisely the system is modeled. Finally, GOAT is fast, being a parallelizable, gradient driven, optimization method.

This theoretical and numerical advance provides a significant step toward the application of numerical optimal control to superconducting quantum computing platforms. The flexibility in pulse description allows the reduction of the number of parameters describing the optimized pulses, reducing the calibration time by an order of magnitude, making it feasible. To our knowledge, this is the first study showing the power of numerical optimal control is applicable to solid state qubits experiments.

\begin{acknowledgments}
S.M. and F.K.W. acknowledge funding from the IARPA through the LogiQ grant No. W911NF-16-1-0114. E.A. acknowledges support from the Alexander von Humboldt Foundation. D.J.T. acknowledges support from the Israel Science Foundation (1094/16) and the German-Israel Foundation for Scientific Research and Development (GIF). F.K.W. acknowledges support from the European Union through the SCALEQIT Project.
\end{acknowledgments}


\bibliography{GOAT_paper_1_bibliography_v2}


\pagebreak
\onecolumngrid

\section{Appendix A - Gradient-free vs. gradient-based search: Nelder-Mead vs. GOAT using L-BFGS}
\label{sec:Appendix-A}

In situations when both gradient-free and gradient-based optimization may be used (e.g. the typical QOC open-loop optimization, with which we are dealing in this work), the critical criteria for deciding whether to make use of one or the other is the efficiency with which one may compute the gradient of the goal function. It is well known that when the gradient can be computed quickly, gradient-based methods, such as L-BFGS, can be orders-of-magnitude quicker than gradient-free methods \cite{nocedal2006no}.

The most commonly used gradient-free optimizer is Nelder-Mead \cite{Nelder-Mead}. We shall therefore compare it to GOAT, which, in this context, can be viewed as a method of computing the goal function gradient, coupled to a standard gradient-based search, L-BFGS \cite{L-BFGS}.

To demonstrate this superiority of GOAT using L-BFGS (GOAT being the method of computing the gradient and L-BFGS the gradient-driven search algorithm) vs. Nelder-Mead, we consider a toy model of a 2 qubit xmon system \cite{Xmon-PRL} (which me model as qutrits), and the resonator is sufficiently detuned to be neglected. The Hamiltonian is

\begin{align}
H_0 = &  g_{12}(b_1^\dagger b_2 + b_1 b_2^\dagger) + \sum_{k=1}^2(\omega_k - \frac{\delta_k}{2})b_k^\dagger b_k+ \frac{\delta_k}{2} b_k^\dagger b_k b_k^\dagger b_k \\
H_l^c = & b_l^\dagger b_l
\label{eq:one-xmon}
\end{align}
where $\omega_k$ is the frequency of the Xmon k and $\delta_k$ is its anharmonicity, $g_{12}$ is the coupling between two neighbouring qutrits. The parametrization used is a 500-slice PWC and the gate generated is a CNOT. Numeric values are $\omega_1=5.23\times2\pi\textrm{ GHz}$, $\omega_2=5.78\times2\pi\textrm{ GHz}$, $\delta_1=-0.220\times2\pi\textrm{ GHz}$, $\delta_2=-0.210\times2\pi\textrm{ GHz}$,  $g_{12}=0.030\times2\pi\textrm{ GHz}$.

\begin{figure}[htbp]
\begin{center}
\includegraphics[scale=0.50]{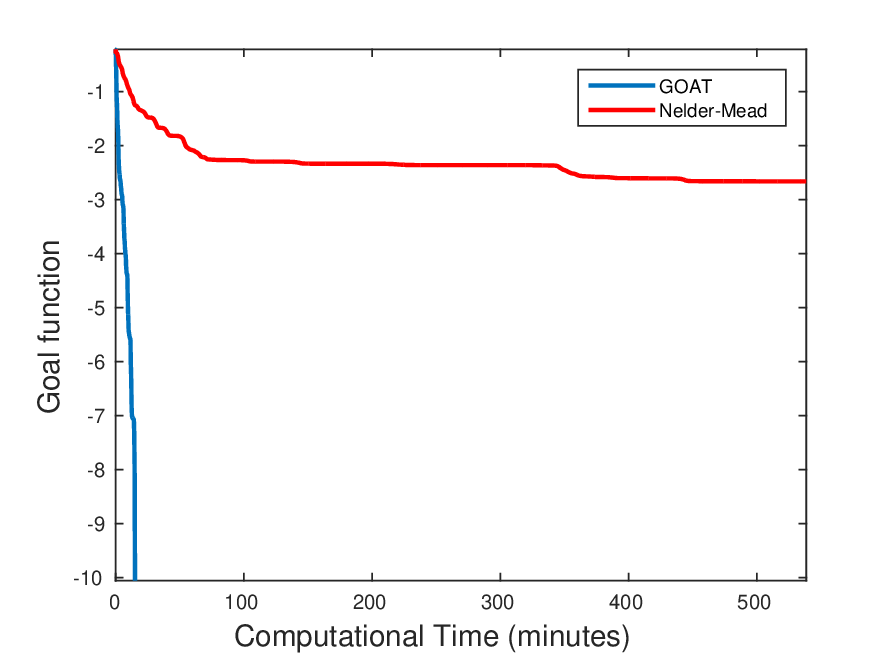}
\end{center}
\caption{Performance of GOAT vs. Nelder-Mead search, as applied to a CNOT gate generation on Xmon system \cite{Xmon-PRL}. Details of the system appear in eq. \ref{eq:one-xmon}. It is evident that GOAT optimization is faster than Nelder-Mead by several orders of magnitude. These results are not surprising, as it is well understood in the computer science community that if one can efficiently compute the gradients of the goal function with respect to the parametrization (as GOAT can), it is far preferable to conduct a gradient-driven search. To contrast, in closed-loop pulse calibration the gradient is not accessible, in which case gradient-free methods are the only available option.}
\label{fig-NM-is-slow}
\end{figure}

Examining Fig. \ref{fig-NM-is-slow} it is clear the Nelder-Mead is not competitive with GOAT for open-loop optimization, as GOAT is capable of efficiently computing and utilizing the gradient of the goal value with respect to the controls, allowing the use of L-BFGS to efficiently search the parameter space. Nelder-Mead, however, struggles when the number of parameters grows. It is important to stress that the results shown in Fig. \ref{fig-NM-is-slow} are universal, with little dependence on system specifics. This is a demonstration of the general attributes of gradient-driven vs. gradient-free search algorithms, when an efficient ay of computing the gradient is available - a fact which is well-known in the general optimization community. A more detailed discussion of the subject is outside the scope of this work.

\pagebreak


\section{Appendix B - Avoiding piecewise constant approximation of control fields}
\label{sec:Appendix-B}

\begin{figure}[htbp]
\begin{center}
\includegraphics[scale=0.50]{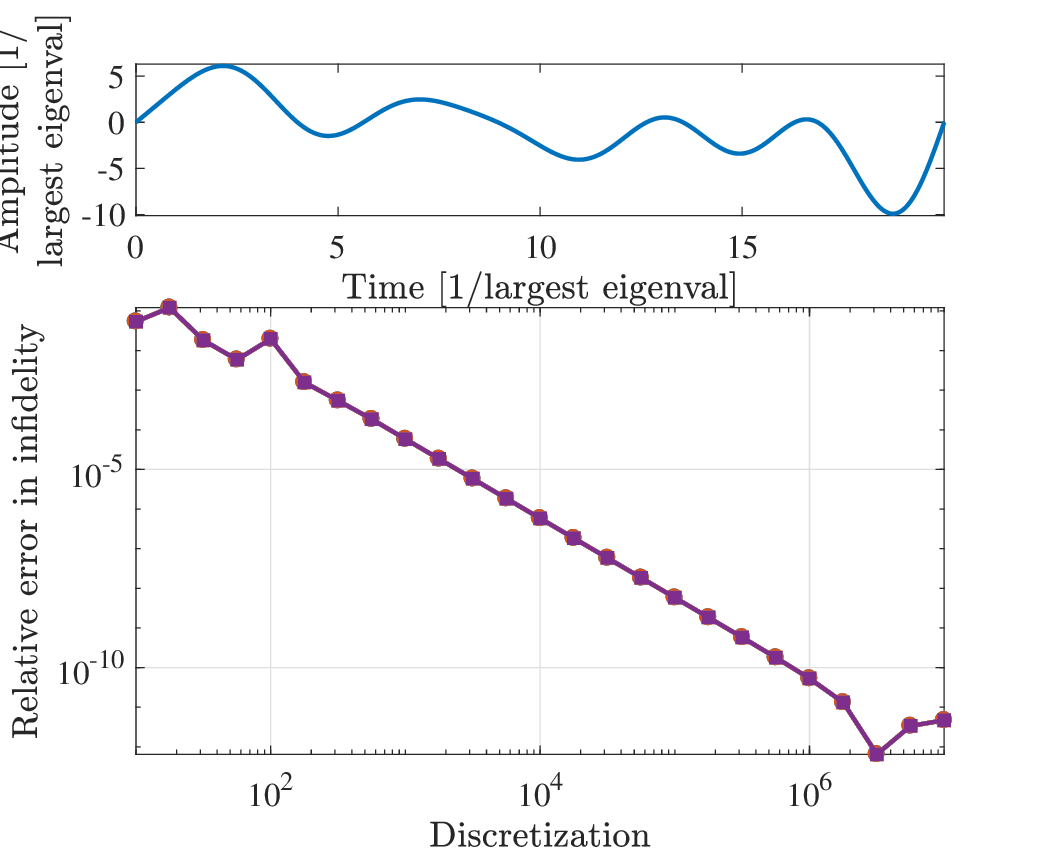}
\end{center}
\caption{Accuracy implications of the piecewise-constant approximation when simulating the dynamics of a 3-spin system with random drift and control Hamiltonians and a smooth control function. PWC samples the control field at middle of equal-width time slices. Top: The smooth drive shape, with 1/time and amplitude specified in the units of the largest magnitude eigenvalue. Bottom: Gate error, comparing the Runge Kutta propagation and the PWC approximation, as a function of the number of time slices used to represent the drive shape, $1-\textrm{Re}\left(\textrm{Tr}\left(U_\textrm{RK}^\dagger U_\textrm{PWC}\right)\right)$. Evidently, in order to achieve high accuracy using the PWC approximation an extraordinarily large number of slices is required.}
\label{fig-PWC-approx}
\end{figure}

Approximating smooth drive shapes by piecewise-constant (PWC) functions introduces inaccuracies into the simulated dynamics \cite{PWC-is-crap-1,PWC-is-crap-2}. This approximation is acceptable if one is seeking low or medium accuracy (infidelity of $10^{-3}$ or below). However, when targeting high fidelities, this approach becomes untenable. To illustrate the point, we simulated the dynamics of a system with a 3-spin random drift Hamiltonian, driven by a second random control Hamiltonian. The control field is parameterized by 11 bandwidth-limited Fourier components, so as to generate a low-frequency, smooth control field. See Fig. \ref{fig-PWC-approx} (top).

We compare the unitary propagator generated by the TDSE $\partial_t U\left(t\right) = \left(H_0 + c\left(t\right) H_c\right) U\left(t\right)$, as computed by a high-accuracy Runge-Kutta 4/5 ODE solver, to integrations of PWC approximations of the control field $c\left(t\right)$ at various resolutions.

In all PWC approximations, the constant Hamiltonian assigned to each slice is taken by sampling $c\left(t\right)$ at the middle of each slice. The error, as a function of the number of pieces used to approximate the smooth drive, is presented in Fig. \ref{fig-PWC-approx} (bottom).

We observe that, for example, more than $\mathcal{O}\left(10^{4}\right)$ time steps are necessary to achieve an error below $10^{-5}$. At such a high step-count, it is numerically advantageous for forgo the PWC approximation and instead fall back on the more accurate Runge-Kutta integrators. Moreover, regardless of questions of numerical expediency in integration, such a representation of the control signal would imply that optimization is carried out in a very high dimensional search space, leading to convergence problems. Even for more sophisticated versions of GRAPE \cite{optim-Skinner}, the exceedingly large Jacobian required to move from the PWC gradient to a more compact parametrization makes the algorithm problematic, as pointed out in \cite{optim-quan-comp1}. Or if the latter approach is not taken, clearly such a huge number of parameters is not calibratable.

\pagebreak

\section{Appendix C - Control Parametrization}
\label{sec:Appendix-C}

The following is a detailed description of the parametrization chain, including both discrete parameter transformations and control function wrappers, used in the flux-tunable coupler \cite{IBM-tunable-coupler} iSWAP optimization described in the main text. We recall the Hamiltonian

\def\arraystretch{1.5}
\begin{equation}
\begin{array}{rcl}
H\left(t\right) & = &  \sum_i \omega_i a_i^\dagger a_i + \omega_{TB}(t)b^\dagger b + g_i(a_i^\dagger b + b^\dagger a_i) - \alpha \left| 2 \right\rangle \left\langle 2\right|_i \, \\
\omega_{TB}\left(t\right) & = & \omega_{\textrm{TB},0}\sqrt[]{|\cos(\pi \Phi/\Phi_0)|} \\
\Phi\left(t\right) & = & \Theta + \delta\left(t\right) \cos(\omega_\Phi t)
\end{array}
\label{eq-H}
\end{equation}
\def\arraystretch{1.0}

Optimization is performed over 18 unconstrained parameters which ultimately describe $\delta\left(t\right)$ using 6 additive frequency components, each with 3 parameters (amplitude, frequency and phase).

The raw optimized parameters are:

\begin{equation}
\begin{array}{rcrlrcrlrcr}
a_{1,1} & = & -0.441194 & \,\,\,\,\,\,\,\, & a_{1,2} & = & 0.412071 & \,\,\,\,\,\,\,\, & a_{1,3} & = & -0.126278\\
a_{2,1} & = & -1.57074 & \,\,\,\,\,\,\,\, & a_{2,2} & = & 0.532019 & \,\,\,\,\,\,\,\, & a_{2,3} & = & 0.134298\\
a_{3,1} & = & 0.43699 & \,\,\,\,\,\,\,\, & a_{3,2} & = & 0.603436 & \,\,\,\,\,\,\,\, & a_{3,3} & = & 0.461675\\
a_{4,1} & = & -1.04331 & \,\,\,\,\,\,\,\, & a_{4,2} & = & 0.490129 & \,\,\,\,\,\,\,\, & a_{4,3} & = & -0.361516\\
a_{5,1} & = & -1.16992 & \,\,\,\,\,\,\,\, & a_{5,2} & = & 0.454957 & \,\,\,\,\,\,\,\, & a_{5,3} & = & 0.91013\\
a_{6,1} & = & 0.81492 & \,\,\,\,\,\,\,\, & a_{6,2} & = & 0.489283 & \,\,\,\,\,\,\,\, & a_{6,3} & = & 0.128118
\end{array}
\end{equation}

The parameters are then linearly rescaled:

\begin{equation}
\begin{array}{rccrrr}
b_{*,1} & = & L\left(a_{*,1},\,-1,1\,\right. & -0.3, & 0.3 & \left.\right)\\
b_{*,2} & = & L\left(a_{*,2},\,-1,1\,\right. & {-2.0944{\times}10^{9}}, & {2.0944{\times}10^{9}} & \left.\right)\\
b_{*,3} & = & L\left(a_{*,3},\,-1,1\,\right. & -3141.59, & 3141.59 & \left.\right)
\end{array}
\end{equation}

where
\begin{equation}
L\left(x,a,b,a_2,b_2\right):=\tfrac{b_2-a_2}{b-a}\left(x-\tfrac{b+a}{2}\right)+\tfrac{b_2+a_2}{2},
\end{equation}

resulting in

\begin{equation}
\begin{array}{rrrrrrrrrrr}
b_{1,1} & = & -0.132358 & \,\,\,\,\,\,\,\, & b_{1,2} & = & {8.6304{\times}10^{8}} & \,\,\,\,\,\,\,\, & b_{1,3} & = & -396.713\\b_{2,1} & = & -0.471222 & \,\,\,\,\,\,\,\, & b_{2,2} & = & {1.11426{\times}10^{9}} & \,\,\,\,\,\,\,\, & b_{2,3} & = & 421.909\\b_{3,1} & = & 0.131097 & \,\,\,\,\,\,\,\, & b_{3,2} & = & {1.26383{\times}10^{9}} & \,\,\,\,\,\,\,\, & b_{3,3} & = & 1450.39\\b_{4,1} & = & -0.312994 & \,\,\,\,\,\,\,\, & b_{4,2} & = & {1.02652{\times}10^{9}} & \,\,\,\,\,\,\,\, & b_{4,3} & = & -1135.74\\b_{5,1} & = & -0.350977 & \,\,\,\,\,\,\,\, & b_{5,2} & = & {9.5286{\times}10^{8}} & \,\,\,\,\,\,\,\, & b_{5,3} & = & 2859.26\\b_{6,1} & = & 0.244476 & \,\,\,\,\,\,\,\, & b_{6,2} & = & {1.02475{\times}10^{9}} & \,\,\,\,\,\,\,\, & b_{6,3} & = & 402.495\tabularnewline
\end{array}
\end{equation}

\setlength\tabcolsep{6pt} 

Next, the parameters are bound,

\begin{equation}
\begin{array}{lll}
c_{*,1} & = &  \mathcal{C}\left(b_{*,1},\,-0.3,0.3\right)  \\
c_{*,2} & = &  \mathcal{C}\left(b_{*,2},\,{-2.0944{\times}10^{9}},{2.0944{\times}10^{9}}\right)  \\
c_{*,3} & = &  b_{*,3} \textrm{(unbounded)}\\
\end{array}
\end{equation}

where the bounding is performed via a scaled sine function:

\begin{equation}
\mathcal{C}\left(x,a,b\right):=\tfrac{b-a}{2}\,\sin\left(\left(x-\tfrac{b+a}{2}\right)/\tfrac{b-a}{2}\right)+\tfrac{b+a}{2}.
\end{equation}

The resultant parameter are

\begin{equation}
\begin{array}{lrrrlrrrlrr}
c_{1,1} & = & -0.128106 & \,\,\,\,\,\,\,\, & c_{1,2} & = & {8.38822{\times}10^{8}} & \,\,\,\,\,\,\,\, & c_{1,3} & = & -396.713\\
c_{2,1} & = & -0.3 & \,\,\,\,\,\,\,\, & c_{2,2} & = & {1.06243{\times}10^{9}} & \,\,\,\,\,\,\,\, & c_{2,3} & = & 421.909\\
c_{3,1} & = & 0.126964 & \,\,\,\,\,\,\,\, & c_{3,2} & = & {1.18852{\times}10^{9}} & \,\,\,\,\,\,\,\, & c_{3,3} & = & 1450.39\\
c_{4,1} & = & -0.259223 & \,\,\,\,\,\,\,\, & c_{4,2} & = & {9.85915{\times}10^{8}} & \,\,\,\,\,\,\,\, & c_{4,3} & = & -1135.74\\
c_{5,1} & = & -0.276216 & \,\,\,\,\,\,\,\, & c_{5,2} & = & {9.20327{\times}10^{8}} & \,\,\,\,\,\,\,\, & c_{5,3} & = & 2859.26\\
c_{6,1} & = & 0.218301 & \,\,\,\,\,\,\,\, & c_{6,2} & = & {9.84352{\times}10^{8}} & \,\,\,\,\,\,\,\, & c_{6,3} & = & 402.495\\
\end{array}
\end{equation}

These rescaled and bounded parameters are then used by the Fourier parametrization

\begin{equation}
d\left(t\right) = \sum_k c_{k,1} \,\sin\left(  c_{k,2}\,t  + c_{k,3}\right)
\end{equation}

Plugging in the transformed optimized parameter values we arrive at

\begin{equation}
\begin{array}{rrrrrr}
c\left(t\right) & = &-0.128106 & \,\sin\left(\right. & {1.33503{\times}10^{8}}\,\,2\pi\,t & \left.-0.277691\,\pi\right)\\
   &  & -0.3 & \,\sin\left(\right. & {1.69091{\times}10^{8}}\,\,2\pi\,t & \left.+0.297666\,\pi\right)\\
   &  & +0.126964 & \,\sin\left(\right. & {1.89158{\times}10^{8}}\,\,2\pi\,t & \left.-0.325411\,\pi\right)\\
   &  & -0.259223 & \,\sin\left(\right. & {1.56913{\times}10^{8}}\,\,2\pi\,t & \left.+0.483735\,\pi\right)\\
   &  & -0.276216 & \,\sin\left(\right. & {1.46475{\times}10^{8}}\,\,2\pi\,t & \left.+0.130343\,\pi\right)\\
   &  & +0.218301 & \,\sin\left(\right. & {1.56664{\times}10^{8}}\,\,2\pi\,t & \left.+0.118224\,\pi\right)\\
\end{array}
\end{equation}

Next, we bind the total additive amplitude and concurrently ensure the controls smoothly go to zero at the beginning and end of the pulse. The smooth edges are enforced by a product of an ascending sigmoid and time-shifted descending sigmoid, and the amplitude bound is enforced by an additional scaled sigmoid. The explicit function is:

\begin{equation}
\delta\left(t\right) = \mathcal{A}\left(d\left(t\right),t/T_{\textrm{final}},\,-0.3,0.3,\,40,0.075\right)
\label{eq-delta-bounded}
\end{equation}

where

\begin{equation}
\begin{array}{lll}
\mathcal{A}\left(f\left(t\right),\tau,\,a,b,\,g,\Delta_{\tau}\right) & := & {\tilde{\mathcal{S}}_{\uparrow}\left(\tau-\Delta_{\tau},g\right)} {\tilde{\mathcal{S}}_{\downarrow}\left(\tau-\left(1-{\Delta_{\tau}}\right),g\right)} \, \mathcal{S}_{\uparrow}\left(-\tfrac{{b}-{a}}{2},\tfrac{{b}-{a}}{2},\,f\left(t\right)\right)
\end{array}
\end{equation}

Here $\tau\in\left[0\ldots 1\right]$ is the scaled time, $a$ and $b$ are the minimal and maximal allowed values, $g$ is the ascent/descent gradient of the window function and $\Delta_{\tau}$ is the width of the ascent/descent. The first sigmoid ensures zero control at $t=0$; the second sigmoid ensures zero control at $t=T_{\textrm{final}}$; the third bounds the amplitude to the $\left[a\ldots b\right]$ range.

The descending [ascending] sigmoid function $\tilde{\mathcal{S}}_{\downarrow}$ [$\tilde{\mathcal{S}}_{\uparrow}$], with value $\tfrac{1}{2}$ at $x=0$ and gradient $g_0$ [$-g_0$], are defined by

\begin{equation}
\begin{array}{lll}
\tilde{\mathcal{S}}_{\downarrow}\left(x,g_{0}\right) & := & 1/\left(1+\exp\left(\left(4\times {{g_{0}}}\right)\times {x}\right)\right) \\
\tilde{\mathcal{S}}_{\uparrow}\left(x,g_{0}\right) & := & 1-\tilde{\mathcal{S}}_{\downarrow}\left(x,g_{0}\right).
\end{array}
\end{equation}

The scaled ascending sigmoid with gradient 1 mid-range is
\begin{equation}
\mathcal{S}_{\uparrow}\left(a,b,\,x\right):=\left(2 \tilde{\mathcal{S}}_{\uparrow}\left(\left(y-\tfrac{{b}+{a}}{2}\right)/\tfrac{{b}-{a}}{2},\tfrac{1}{2}\right) - 1\right) \tfrac{{b}-{a}}{2} + \tfrac{{b}+{a}}{2}
\end{equation}

The bounded $\delta\left(t\right)$ of eq. \ref{eq-delta-bounded} serves as input to the arbitrary waveform generator, where a carrier and bias are added
\begin{equation}
\Phi\left(t\right) = \mathcal{G}\left(\delta\left(t\right),t,\,-0.108,{5.34448{\times}10^{9}}\right)
\end{equation}
where
\begin{equation}
\mathcal{G}\left(f\left(t\right),t,\,\Theta,\omega_{\Phi}\right) := \Theta + \cos\left(\omega_{\Phi}t\right) f\left(t\right).
\end{equation}

Finally, the resultant flux takes the following form as the pre-factor of the control Hamiltonian $b^\dagger b$,
\begin{equation}
\omega_{\textrm{TB}}\left(t\right) = \mathcal{W}\left(\Phi\left(t\right),t,\,{4.67783{\times}10^{10}}\right)
\end{equation}
where
\begin{equation}
\mathcal{W}\left(\Phi\left(t\right),t,\,\omega_{\textrm{TB}_0}\right) := \omega_{\textrm{TB}_0}\sqrt{\left|\cos\left(\pi  \Phi\left(t\right)\right)\right|}
\end{equation}

The gradient of the above function compositions is used by the optimization algorithm.

Should one wishes, it is possible to accurately simulate the effects of an Arbitrary Waveform Generators (AWG) on the generated signal by
\begin{itemize}
  \item Sampling the smooth control ansatz at specific times ($1.2$GHz is a commonly used sampling rate).
  \item Generating PWC output from the AWG.
  \item Filtering the PWC output with a low-pass filter, which is inherent to all AWGs ($0.3$Ghz is again a commonly used bandwidth).
\end{itemize}

Adapting GOAT algorithm to the above model would require
\begin{itemize}
  \item Generating the Jacobians required for the chain rule describing the discrete sampling $\partial_{\bar{\alpha}}c_k\left(\bar{\alpha},t\right)$.
  \item The transfer functions will be modeled by combining the equations of motion of the filter's output and its internal state (e.g. the equations of motion for charge and current of an RLC bandpass-filter) with the propagator equation of motion, $\partial_tU\left(\bar{\alpha},t\right)$. If one derives the entire set with respect to $\bar{\alpha}$, one arrives at an expanded version of eq. (7) of the main text; see also \cite{hincks2015controlling}.
\end{itemize}

However, the above highly precise modeling of an AWG is not necessary in this case, as the produced pulses are already bandwidth limited to the capabilities of typical AWGs.

\bigskip
\bigskip


\section{Appendix D - Local phase insensitive goal function}
\label{sec:Appendix-D}

Local (single qubit) $Z$ rotations can be accounted for in software by noting the logical rotation of the $x-y$ plane. Therefore, we aim to  generate the iSWAP gate modulo $Z$ rotations. The following goal function is insensitive to such operations

\begin{equation}
g = 1 - \frac{1}{d} \textrm{Tr}\left( |U_{goal}^\dagger U(t_f)|^2\right).
\end{equation}
The gradient is
\begin{equation}
\partial_\alpha g = - \frac{2}{d}\textrm{Tr}\left( |U_{goal}|\textrm{Re}(\partial_\alpha U(t_f) .* U(t_f)^*)\right)
\end{equation}
where $.*$ denotes element-wise multiplication. Notice that both the goal function and the gradient are defined element-wise.

This approach can be generalized to any target gate, and will be covered in more detail in forthcoming work.


\end{document}